# ANTICORRELATIONS, SUBBROWNIAN STOCHASTIC DRIFT, AND 1/F-LIKE SPECTRA IN STABLE FINANCIAL SYSTEMS

K.Staliunas

E-mail: Kestutis.Staliunas@PTB.DE

**Abstract**

Statistic dynamics of financial systems is investigated, basing on a model of randomly coupled equation system driven by stochastic Langevin force. It is found that in stable regime the noise power spectrum of the system is of $1/w^a$ form, with the exponent $a = 3/2$ in case of Hermitian coupling matrices, or slightly larger (up to $a = 3$) in nonhermitian case. In all cases it leads to negative autocorrelation function of the increments of the variables (prices, exchange rates), and to subbrownian stochastic drift of the variables. The model can be generalized to arbitrary stable, driven by noise system of randomly coupled components.

# I. Introduction

The analysis is initiated by observations that the temporal autocorrelation functions of increments of variables $C(t) = <dx(t) \cdot dx(t-t)>$ in liquid finance markets are universally nonzero for small times $t$ [1]. ($dx(t)$ is the increment of the variable $x(t)$ which corresponds to price, exchange rate and similar.) Most likely the autocorrelation function follows the power law $C(t) \propto -t^g$ for temporal delays $t$ approximately until one hour, with the exponent $g$ of the order of unity. The aim of this paper is find an explanation of this nonzero autocorrelation function, and to calculate its particular form.

The negative autocorrelations can be related with a subbrownian stochastic drift of the corresponding variable $x(t)$. Under "subbrownian stochastic drift" is assumed that the variance of stochastically drifting variable increases as $<(x(t) - x(0))^2> \propto t^b$ with increasing time $t$, with $b < 1$. For comparison, the variance of a position of Brownian particle increases linearly with time $b = 1$, following the well known Wiener law of stochastic drift.

The negative autocorrelations can be also related with 1/f - like form of the power spectra: $S(w) \propto w^{-a}$ with $a < 2$.

Recent investigations have shown that stochastic drift of invariant variables (e.g. the phase of the order parameter, the position of a localized structure or of a vortex) in stable, spatially extended, and driven by noise nonlinear systems is in general subbrownian, and that the power spectra are 1/f - like, with the exponent $a$ explicitly dependent on the dimensionality of the space [2]. A crucial point in the study [2] is that the spatially extended system has infinitely many degrees of freedoms. Summation over degrees of freedom (or integration in the limit of continuum) allowed to obtain above listed peculiarities of stochastic distributions (subbrownian drift, 1/f-like spectra), absent in the systems with small number of degrees of freedom. Motivated by the results from [2] here a similar analysis is performed for stable financial markets where the number of traders is usually very large. In the limit of infinity large number of traders one can consider the system as a continuum, thus the dynamics in such system may resemble that of continuous spatially extended systems.

Summarizing the analysis used in [2]:

1) one starts with a nonlinear evolution equation:

$$\frac{d\mathbf{A}}{dt} = \mathbf{NL} \cdot \mathbf{A} + \tilde{\mathbf{A}}(\mathbf{r},t) \qquad (1)$$

for the temporal evolution of the state vector (order parameter) of the system: $\mathbf{A}(t) = (A_1(t), A_2(t), ..., A_j(t), ..., A_n(t))^T$. $\mathbf{NL}$ is a (deterministic) nonlinear operator acting on the order parameter, and $\tilde{\mathbf{A}}(t)$ is the vector of $d$ - correlated noise terms $\tilde{\mathbf{A}}(t) = (\Gamma_1(t), \Gamma_2(t), ..., \Gamma_j(t), ..., \Gamma_n(t))^T$, of temperature $T$: $<\Gamma_j(t_1) \cdot \Gamma_k(t_2)> = 2T \cdot d_{jk} d(t_1 - t_2)$.

2) one linearizes around the stationary solution $\mathbf{A}_0$ of the equation (1) without the noise term: $\mathbf{A}(t) = \mathbf{A}_0 + \mathbf{a}(t)$, and obtains the linear equation system:

$$\frac{d\mathbf{a}(t)}{dt} = \mathbf{L} \cdot \mathbf{a}(t) \qquad (2)$$

Rewriting (2) in a new basis $\mathbf{a}(t) \to \mathbf{b}(t)$ corresponding to the eigenvectors of the matrix $\mathbf{L}$, system) one obtains:

$$\frac{d\mathbf{b}(t)}{dt} = \ddot{\mathbf{E}} \cdot \mathbf{b}(t) + \tilde{\mathbf{A}}'(t) \qquad (2)$$

where $\ddot{\mathbf{E}}$ is a diagonal matrix consisting of the eigenvalues of the matrix $\mathbf{L}$, and $\tilde{\mathbf{A}}'(t)$ is the noise vector in the new coordinates.

3) one calculates the power spectra of the normal modes $\mathbf{b}(t)$ as driven by noise. After Fourier transform $\mathbf{b}(t) \to \mathbf{b}(w) = \int \mathbf{b}(t)\exp(iwt)dt$ one obtains the expressions for spectra of each perturbation mode: $b_j(w) = \Gamma'_j(w)/(iw + l_j)$.

4) one makes summation (or integrating in the limit of continuum) of the spectra of normal modes to obtain the total power spectrum of the system: $S(w) = \sum_j S_j(w) = \sum_j |b_j(w)|^2 = \int |b_j(w)|^2 dj$.

We use essentially the same procedure for analysis of stable liquid financial markets (or at least of stable periods of generally unstable markets). Without discussing the particular financial model we assume very generally that a large number of traders are nonlinearly coupled one with another, which can be described by operator $\mathbf{NL}$. The operator $\mathbf{NL}$ includes all the trading strategies of traders one against another. It is also assumed that the financial market is subjected by random $d$ - correlated perturbations, corresponding to political events and similar. It corresponds to external noise of particular temperature $T$. Then the system under investigation can be formally described by (1).

Basing on a general model (1), and using some additional assumptions, a concrete form of equations can be written. We however will not try to guess the form of nonlinear operator $\mathbf{NL}$ in (1), but, assuming that the system is stable, will try to guess the form of linear matrix $\mathbf{L}$ in (2). Indeed, if the system (1) has a stationary solution, then the linearization of corresponding nonlinear operator $\mathbf{NL}$ is in principle possible in the vicinity of the this solution: $\mathbf{A}(t) = \mathbf{A_0} + \mathbf{a}(t)$, where $\mathbf{a}(t) = (a_1(t), a_2(t),...,a_j(t),...,a_n(t))^T$ stands for the perturbation of state vector of the system with $n$ is the number of the traders very large. Then (2) is also possible. Since the trading strategies of different traders may be very different, then it is plausible to assume that the coupling matrix $\mathbf{L}$ is a random matrix. And the stability assumption imposes a condition on the coupling matrix $\mathbf{L}$, that it must be negative definite.

In this sense the very general assumptions, that: 1) the system is stable; 2) the traders have many different strategies; 3) the system is subjected to external random perturbations, lead to the model (2). Differently from the "deterministic" case (e.g. complex Ginzburg-Landau equation with noise as investigated in [2]), where the linear coupling matrix $\mathbf{L}$ is derived from the known nonlinear model, for financial markets the matrix $\mathbf{L}$ should be "guessed" or introduced phenomenologically, due to the lack of the knowledge on the underlying processes. The most natural guess is that the matrix $\mathbf{L}$ is a random negative definite matrix.

The random matrix $\mathbf{L}$ is then to be diagonalized, which generates Lyapunov exponents distributed over the complex $l$-plane. Then the power spectra of the corresponding normal modes are to be calculated as driven by external noise. The summation over the spectra of the modes (or the integration in the limit of continuum, when the number of traders tend to infinity) leads eventually to the power spectrum of a particular variable of the system.

## II. Choice of random matrix;

The random matrix $\mathbf{L}$, as mentioned above, should be negative definite. The form of the matrix, under the restriction of negative definite must be as simple and as natural as possible. Two models for the random matrix $\mathbf{L}$ were chosen:

1) As the most natural choice of a real-valued negative number is $-x^2$ then the coupling matrix is constructed as $\mathbf{L} = -\mathbf{M} \cdot \mathbf{M}^T$, where $\mathbf{M}$ is square matrix of normally-distributed real numbers, and $\mathbf{M}^T$ is its transpose matrix. In this case $\mathbf{L}$ is Hermitian, symmetric, negatively definite matrix, thus all its eigenvalues are real and negative. This choice leads to analytic results, however the situation does not precisely corresponds to the reality in financial markets. The symmetry of the coupling matrix means reciprocallity of the strategies of one trader towards another, which is perhaps not always the case in the real financial markets.

2) As the most natural form of a complex number with negative real part is $-x_1^2 + icx_2$ then the coupling matrix is constructed as $\mathbf{L} = -\mathbf{M}_1 \cdot \mathbf{M}_1^T + c(\mathbf{M}_2 - \mathbf{M}_2^T)$. In this case $\mathbf{L}$ is nonhermitian, negatively definite matrix, thus its eigenvalues are complex-valued numbers with negative real part. No analytical results were obtained in this case, and only numerical analysis was performed.

## III. Hermitian, negatively definite matrix
### 1) Power spectra

The density of the eigenvalues of Hermitian random matrix $\mathbf{L}$ is:

$$\rho(\lambda) = \frac{1}{2\pi\sigma^2} \sqrt{\frac{4\sigma^2 - \lambda}{\lambda}} \tag{4}$$

with $\sigma^2/n$ being the variance of the elements of matrix $\mathbf{M}$ of size $(n \times n)$.

The response of each perturbation mode to external noise, as derived directly from (3) is:

$$b_j(\omega) = \frac{\Gamma_j(\omega)}{i\omega + \lambda_j} \tag{5}$$

The power spectrum of each perturbation mode (assuming delta correlated noise of temperature T: $\langle \Gamma_j(\omega) \cdot \Gamma_k^*(\omega) \rangle = T\delta_{j,k}$) is:

$$S_j(\omega) = |b_j(\omega)|^2 = \frac{T}{\omega^2 + \lambda_j^2} \tag{6}$$

The power spectrum of the system is a sum of power spectra of each normal mode (6), $S(\omega) = \sum_j S_j(\omega)$, which in the limit of infinitely large number of traders can be replaced by the integral of power spectrum (6) over eigenvalues, accounting for the density function (4):

$$S(w) = \int s(l) \cdot S_l(w) \cdot dl = \frac{T}{w^{3/2}s} \left(1 + (4r^2/w)^{-2}\right)^{1/4} \sin\left(\frac{arctan(4r^2/w)}{2}\right) \qquad (7)$$

which in the limit of small frequencies $w \ll 4s^2$ simplifies to:

$$S(w) = \frac{T}{w^{3/2}s\sqrt{2}} \qquad (8)$$

which means the 1/f-like power spectra with the exponent $a = 3/2$.

A numerical check of the spectrum (8) was performed. The matrices **M** of normally distributed real random numbers of size $(1000 \times 1000)$ and with $s = 1$ were generated numerically. The density of the eigenvalues of the corresponding Hermitian matrix is shown in Fig.1.a (as averaged over 10 realizations). The corresponding spectra in log-log scale, as calculated by summation the response functions (6) over the normal modes is shown in Fig.1.b. The region characterized by 3/2 slope, as extending to $w_{max}$ approximately of order of unity is clearly visible in accordance with analytic results. For large frequencies $w > 1$ the slope is 2, indicating that the usual random Markov process takes place for very short times.

**2) Stochastic drift**

The integral of the 1/f-like power spectrum diverges in the limit of small frequency, which means that the variance of the increment of the variable grows to infinity for large times. The variance of the variable $\langle \Delta x(t)^2 \rangle = \langle (x(t) - x(0))^2 \rangle$ can be calculated from the power spectra:

$\langle \Delta x(t)^2 \rangle \approx \int_{w_{min}}^{\infty} S(w) dw$, where $w_{min} \approx 2p/t$ is the lower cut-off boundary of the power spectrum (as follows from the Parseval theorem). Thus the variance of the increment obeys $\langle \Delta x(t)^2 \rangle \propto t^{a-1}$ for the power spectra $1/w^a$. This generalizes the well known Wiener law for stochastic drift, stating that the variance grows linearly with time (or equivalently the root mean wandering obeys the square root $\sqrt{\langle \Delta x(t)^2 \rangle} \propto t^{1/2}$ law). From the results above it follows that the stochastic drift in financial time series should be weaker than in Brownian diffusion. In particular the variance in the model described by Hermitian random coupling matrix, ($a = 3/2$) increases as $\langle \Delta x(t)^2 \rangle \propto t^{1/2}$.

**3) Autocorrelation functions**

The autocorrelation function of the increments can be calculated from the power spectra applying the Wiener-Kinchin theorem: $K(t) = \int S'(w) \exp(-iwt) dw$, here $S'(w)$ stands for the power spectrum of the increments related with the power spectrum of original variable by $S'(w) = w^2 S(w)$. The flat power spectrum of increments $S'(w) = const$, or equivalently the $S(w) \propto 1/w^2$ power spectrum of variable, leads to $d$ - correlations as it is well known for Wiener process (Brownian drift). The power spectra $S(w) \propto 1/w^a$ as calculated above in general with $a < 2$, leads to the autocorrelation function of the increments $K(t) \propto -t^{-g}$, with $g = 3 - a$. In particular the autocorrelation function of the increments described by Hermitian random coupling matrix follows $K(t) \propto -t^{-3/2}$ dependence.

The autocorrelation function as obtained numerically is shown in Fig.1.c.

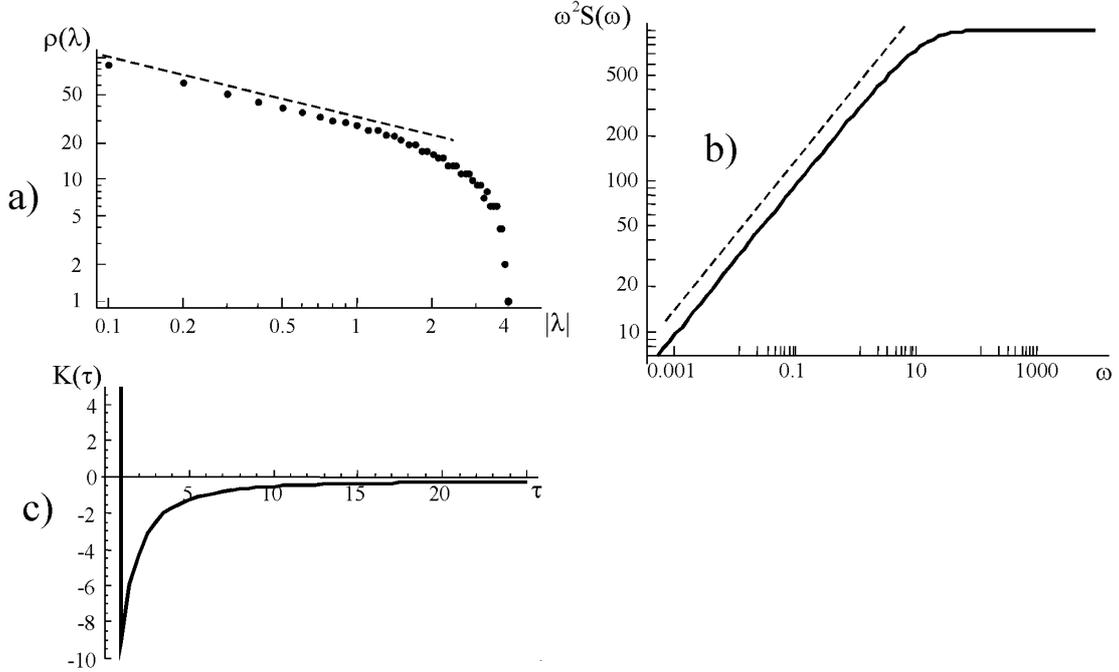

**Fig.1**. a) Distribution of the eigenvalues of numerically generated Hermitian negatively definite random matrices of size ($1000 \times 1000$) in log-log scale, as averaged over 10 realisations. The dashed line indicates the slope of $-1/2$; b) the normalized power spectrum (multiplied $w^2$), corresponding to power spectrum of the of the increments of variables. The dashed line indicates the slope of $1/2$; c) autocorrelation function of the increments of variables.

**IV. Nonhermitian, negatively definite matrix**

This case was investigated numerically only. A typical realization (the Lyapunov exponents in the $l$ complex plane) is shown in Fig.2.a, from numerically generated matrices $\mathbf{M}_1$ and $\mathbf{M}_2$ of normally distributed random numbers of size ($1000 \times 1000$) and with $s = 1$. The distribution of the real part of the eigenvalues approximately follow the $l^{-1/2}$ law, like in the case of Hermitian matrices. The summation was performed taking into account the following response function:

$$S_j(w) = |b_j(w)|^2 = \frac{T}{(w - l_{j,\text{Im}})^2 + l_{j,\text{Re}}^2} \tag{9}$$

which lead to the power spectrum as depicted in Fig.2.b. The regions of linear slope in the log-log scale were obtained. The slope $a$ depends on the nonreciprocality parameter $c$, and increases from $a_{min} = 1.5$ at the Hermitian limit ($c = 0$), until $a_{max} = 1.666...$ for increasing values of $c$. Like in Hermitian case the power dependence extends over several decades until the frequency of order of unity. For large frequencies the dependence with $a = 2$ is obtained.

Differently from Hermitian case, the saturation of the spectra for small frequencies is observed. It is attributed to the depopulating the regions of $l$ complex plane around the axis $l_{Re} = 0$ (corresponding to of "unlocking" of the slow normal modes in general in nonhermitian cases).

The variance of stochastically drifting variable was not investigated numerically. As follows from the discussion stochastic drift should obey $\langle \Delta x(t)^2 \rangle \propto t^b$ dependence, with $b = a - 1$ varying from 0.5 at the Hermitian limit, to 0.666.

The autocorrelation function of the increments was calculated numerically, and it was checked, that it follows precisely the same $K(t) \propto -t^{-g}$ dependence, with $g = 3 - a$ varying from 1.5 at the Hermitian limit, until 1.333... Visually, the autocorrelation function does not differ much from that in Hermitian limit (shown in Fig.1.a). (The differences are visible in log-log plot).

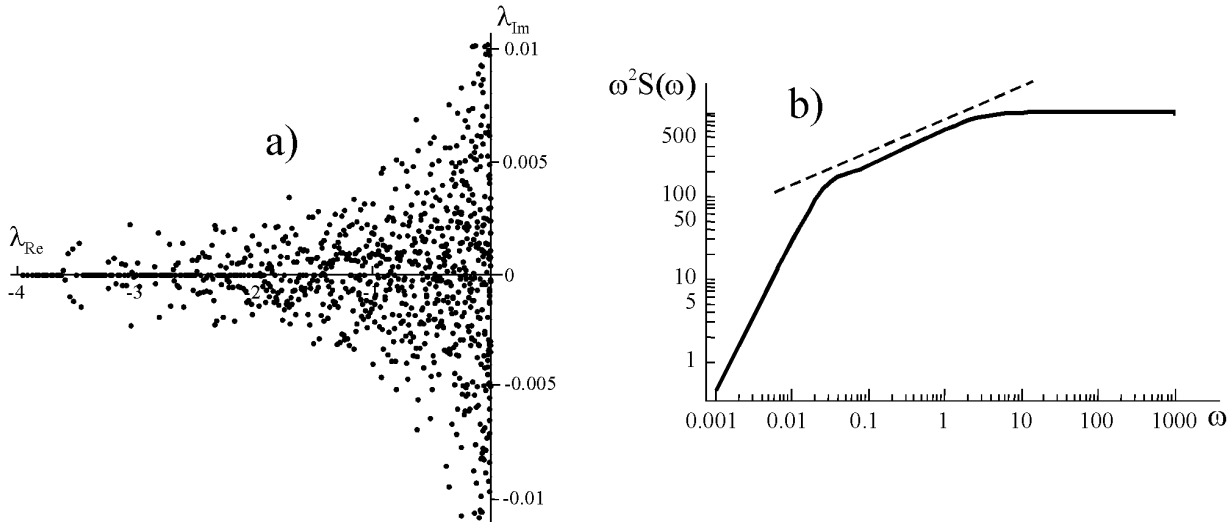

**Fig.2.** a) Distribution of eigenvalues of numerically generated nonhermitian negatively definite random matrices of size ($1000\times1000$), with $s = 1$, and $c = 1$ on the complex $l$ - plane, as averaged over 10 realisations. b) corresponding power spectrum of the increments of variables. The dashed line indicates the slope of $0.4$.

## V. Conclusions

The stochastic properties of the financial time series were calculated using a very general model, assuming that: 1) large number of traders having different strategies is present; 2) external noise is present; 3) system is stable. The calculations lead to 1/f-like power spectra, to negative autocorrelations of increments, and to subbrownian stochastic drift of the variables.

The correspondence with historical financial time series [1] (e.g. high frequency currency exchange rates) is very good for small times (until 1 hour). The correspondence is even better using the filtered time series (cutting out the high volatility periods from the data) [3].

Our model, however ceases to work for large times (small frequencies), and during the periods of large volatility. For large times the autocorrelations vanishes in real markets. Seemingly our model is no more valid for large times, since the traders adapt their strategies depending on the results on large time scales (more than one hour), thus the nonlinear coupling operator becomes nonstationary and simple linearization is no more possible.

During the large volatility periods the system becomes unstable, and, speaking in terms of our model, a part of Lyapunov exponents become positive real part. Therefore our linear model is also no more valid. Moreover, during the large volatility periods the kind of itinerant behaviour is seemingly present, causing the state vector of the system to jump between different fixed points. Such behaviour leads to so called "fat tails" of the probability distributions. In our linear model, where all the dynamics is limited to the vicinity of only one stationary point the statistical distributions of the increment of the variable remains Gaussian. Preliminary the extension of the model to the case of several stable points leads to "fat tail" statistical distributions of the increments, and is the object of the present study.

The results from the paper apply not only to the financial markets, but in general to every noise driven system of randomly coupled components.